\documentclass[page-classic]{epl2} 

\usepackage{graphicx} 

\title{Restricted random walk model as a new testing ground for the applicability 
of $q$-statistics}

\author{Ugur Tirnakli\inst{1} \and Henrik Jeldtoft Jensen\inst{2} \and Constantino Tsallis\inst{3}}
\shortauthor{Ugur Tirnakli \etal}

\institute{
  \inst{1} Department of Physics, Faculty of Science, Ege University, 35100 Izmir, Turkey\\
  \inst{2} Complexity \& Networks Group and Department of Mathematics, Imperial College London, 
South Kensington Campus, London SW7 2AZ, UK\\
  \inst{3} Centro Brasileiro de Pesquisas F\'\i sicas and National Institute of Science and 
Technology for Complex Systems, Rua Dr. Xavier Sigaud 150, 22290-180 Rio de Janeiro, Brazil\\
and\\
Santa Fe Institute, 1399 Hyde Park Road, Santa Fe, NM 87501, USA
}
\pacs{05.20.-y}{Classical statistical mechanics}
\pacs{05.40.Fb}{Random walks and Levy flights}
\pacs{02.60.Cb}{Numerical simulation; solution of equations}

\abstract{We present exact results obtained from Master Equations for the probability function 
$P(y,T)$ of sums $y=\sum_{t=1}^T x_t$ of the positions $x_t$ of a discrete random walker 
restricted to the set of integers between $-L$ and $L$. We study the asymptotic properties 
for large values of $L$ and $T$. For a set of position dependent transition probabilities 
the functional form of $P(y,T)$ is with very high precision represented by $q$-Gaussians 
when $T$ assumes a certain value $T^*\propto L^2$. The domain of $y$ values for which the 
$q$-Gaussian apply diverges with $L$. The fit to a $q$-Gaussian remains of very high 
quality even when the exponent  $a$ of the transition probability $g(x)=|x/L|^a+p$ 
with $0<p<<1$ is different from 1, although weak, but essential, deviation from the 
$q$-Gaussian does occur for $a\neq1$. To assess the role of correlations we compare 
the $T$ dependence of $P(y,T)$ for the restricted random walker case with the 
equivalent dependence for a sum $y$ of uncorrelated variables $x$ each distributed 
according to $1/g(x)$.  }

\begin{document}

\maketitle

\section{Introduction}\label{intro}
The central limit theorem states that appropriately scaled sums of independent random variables will   be distributed according to a Gaussian \cite{vKa,khinchin}. The random walker is the prototype example of a stochastic Gaussian process \cite{feller,reif}. The standard random walker is characterized by transition constant probabilities, which are independent of position and time. Here we point out that for a certain class of position dependent transition probabilities correlations arise, which leads to deviations away from Gaussian behavior. There exist already a large amount of 
evidence, which points to $q$-Gaussians as the relevant high quality approximates for the functional form for the distribution function in a range of cases where correlations play an essential r{\^o}le \cite{tsa3}. 
The evidence for the relevance of the $q$-Gaussian is however often derived from numerical 
experiments in which fluctuations limits the accuracy and therefore the precision of the 
fit to the $q$-Gaussian form (analytical exceptions to this frequent 
difficulty can be found in \cite{rodriguezstefan}). Moreover we believe the random walk example we discuss here to be highly generic. It is related to e.g. particles moving in a confining potential or to more branching processes subject to resource limitations. 

Here we present an investigation of the sum 
$y=\sum_{t=1}^Tx_t$ of positions $x_t$ passed through by a Restricted  Random Walker (RRW). 
The underlying stochastic process is sufficiently simple to allow exact numerical solution 
of the Master Equation (ME) for the probability distribution $P(y,T)$. This ensures a very 
high precision fit to the $q$-Gaussian form and thereby a very accurate determination of the 
relevant parameters. We find that a broad range of {\em transition} probabilities for the 
random walker leads to $q$-Gaussians with $q$ parameters depending on transition 
probabilities. Since the ME can be easily handled in exact numerically form, the RRW model 
is an excellent laboratory for understanding the conditions under which sums of correlated 
random variables are distributed as $q$-Gaussians:
\begin{equation}
P(y) =  \left\{ \begin{array}{ll}
 P(0)\left[1-\beta (1-q) y^2\right]^{\frac{1}{1-q}} &\mbox{for $\beta (1-q) y^2<1$}\\
     0 & \mbox{otherwise}
     \end{array}
     \right.
\label{qGauss}
\end{equation}
where $q<3$ and $\beta>0$ are parameters (for $q\geq 3$ 
normalizability is lost). 
As $q\rightarrow 1$ the function $P(y)$ approaches the Gaussian.

\section{Restricted Random Walk Model}
We consider a one dimensional symmetric random 
walker confined to the integers 
between $-L$ and $L$. The motion of the walker is controled by the following time 
evolution 
 \begin{equation}
x_{t+1} =\left\{\begin{array}{ll}
                              x_{t}+1 &\mbox{with probability $g(x)/2$}\\
                              x_{t}-1 &\mbox{with probability $g(x)/2$}\\
                                x_t &\mbox{with probability $1-g(x)$.}
                                \end{array}
                                \right .
\end{equation}
We concentrate on the following form 
\begin{equation}
g(x)=\min\left\{\left|\frac{x}{L}\right|^a+p,1\right\},
\end{equation}
with reflective boundary conditions: If $x_{t+1}>L$ $(<-L)$ we let 
$x_{t+1}\mapsto x_{t+1}-1$ $(+1)$. We find numerically that the first return time 
(defined as the time elapsed until the walker, who leaves its $x=0$ position, 
returns to the zero position again, and note we do not include walkers who remain 
at $x=0$ for all times) distribution for these RRW behaves asymptotically like 
$P(T)\sim T^{-\tau}$, with $\tau=2$, {\em i.e. different} from the exponent 
$\tau=3/2$ for ordinary RW. 
We study the sum $y=\sum_{t=1}^Tx_t$ in the limit $p\rightarrow 0$ for values of the 
exponent $a=0.75$, $1$ and $1.25$.
For $p\rightarrow 1$ and $L\rightarrow\infty$ the process reduces to the ordinary 
random walk.
 
\begin{figure}
\includegraphics*[scale=0.25]{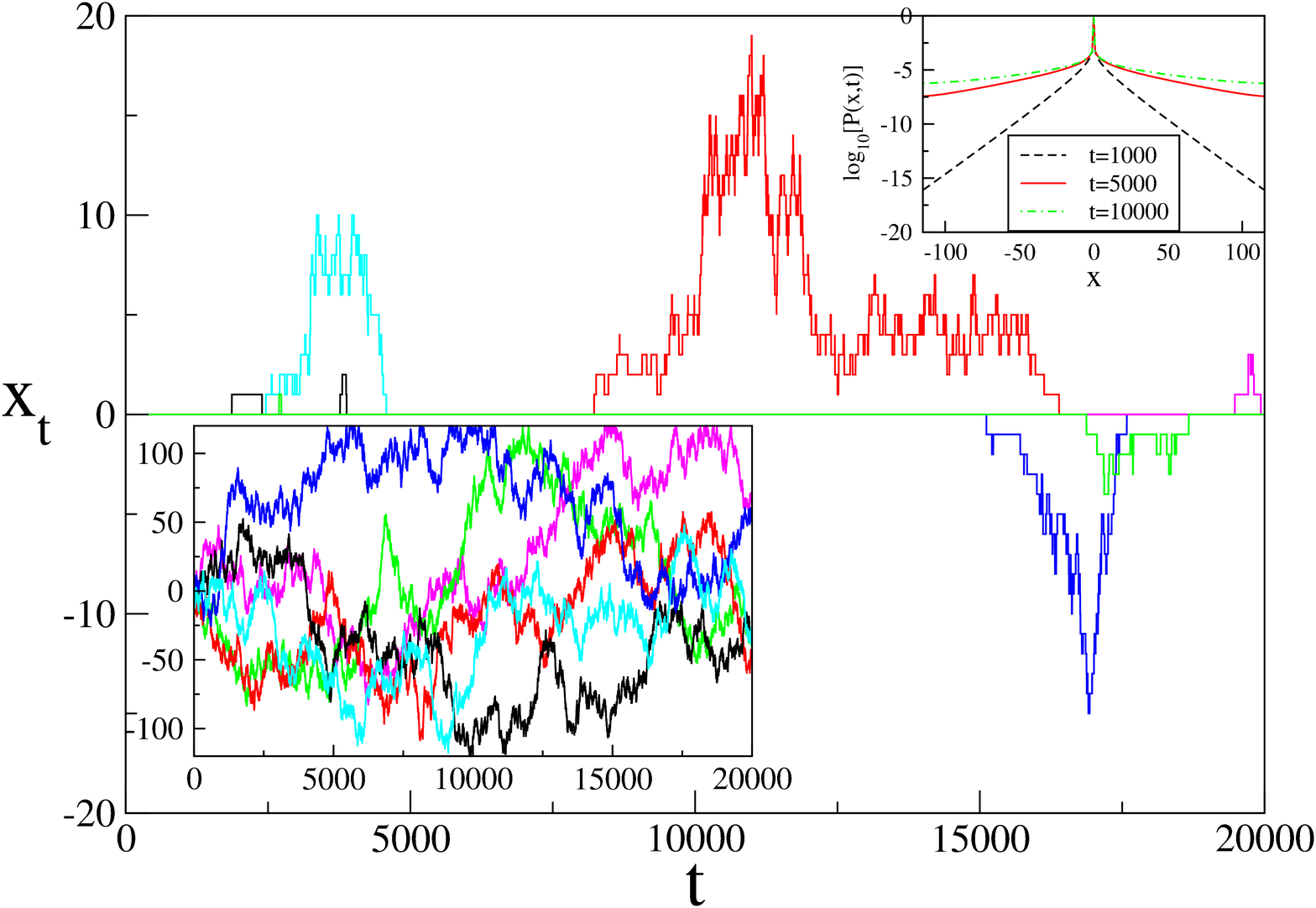}
\caption{Six representative trajectories for $L=120$. 
Main panel: The restricted RW model ($p=5\cdot 10^{-6}$). 
Lower Inset: The standard RW model ($p=1$). 
Non-ergodic behavior of the restricted RW model can easily be seen. Upper Inset: The time evolution of $P(x,t)$ at three different $t$ values ($t$ = 1000, 5000 and 10000 from bottom to top).}
\label{xvst}
\end{figure}

The highly restrictive nature of the RRW is clearly seen from the 6 trajectories shown 
in Fig.\ref{xvst} in the case $L=120$  (with $a=1$ and $p=5\cdot 10^{-6}$). For comparison 
we present the trajectories of an ordinary random walk on $x_{t+1}=x_t\pm1$ with 
probability $1/2$ and $x_t$ confined to $-L,-L+1,..,L-1,L$. The figure shows that the 
vanishing transition probability $g(x)$ near $x=0$ makes the RRW non-ergodic leaving most 
of the phase space empty. It is straight forward to derive a Master Equation for the distribution $P(x,t)$ 
\begin{equation}
P_X(x,t+1)=P_X(x,t)+\frac{1}{2}g(x-1)P_X(x-1,t) +\frac{1}{2}g(x+1)P_X(x+1,t)-g(x)P_X(x,t)
\end{equation}
subject to the appropriate boundary conditions at $|X|=L$. The insert in Fig. \ref{xvst} exhibits a solution of the equation. We will discuss $P(x,t)$ in a more extended publication, here we now turn our attention to another distribution.

  Let $P(y,x,T)$ denote the  probability that $\sum_{t=1}^T x_t=y$ and $x_{T}=x$. The time 
  evolution of this simultaneous probability is controlled by the following ME   
 \begin{eqnarray}
&&P(y,x;t+1)=P(y,x;t)+
\sum_{\Delta\in\{-1,0,1\}} \nonumber\\
&&[W(y,x;y-(x-\Delta),x-\Delta)P(y-(x-\Delta),x-\Delta;t))\nonumber\\
&& -W(y+x+\Delta,x+\Delta;y,x)P(y,x;t)].
\end{eqnarray}
 The transition probabilities $W$ only depend on $x$ and $\Delta$. We have 
\begin{eqnarray}
W(y,x;y-(x-\Delta),x-\Delta)= w(x-\Delta,\Delta)\nonumber
\end{eqnarray}
 and 
\begin{eqnarray} 
W(y+x+\Delta,x+\Delta;y,x)=w(x,\Delta)\nonumber
\end{eqnarray}
 where
\begin{equation}
w(z,\Delta)=\left\{ \begin{array}{ll}
g(z)/2 & \mbox{if $\Delta=\pm 1 $}\\
1-g(z) & \mbox{if $\Delta=0$}
\end{array}
\right .
\end{equation}
By substituting $g(x)$ we obtain the following simple equation
\begin{eqnarray}
&&P(y,x;T+1)=\frac{1}{2}[g(x+1)P(y-(x+1),x+1;T)\nonumber\\
&&+g(x-1)P(y-(x-1),x-1;T)]\nonumber \\&&+(1-g(x))P(y-x,x;T)
\end{eqnarray}
The relevant boundary conditions are straightforward but lengthy to write down. 

\begin{figure}
\includegraphics*[scale=0.5]{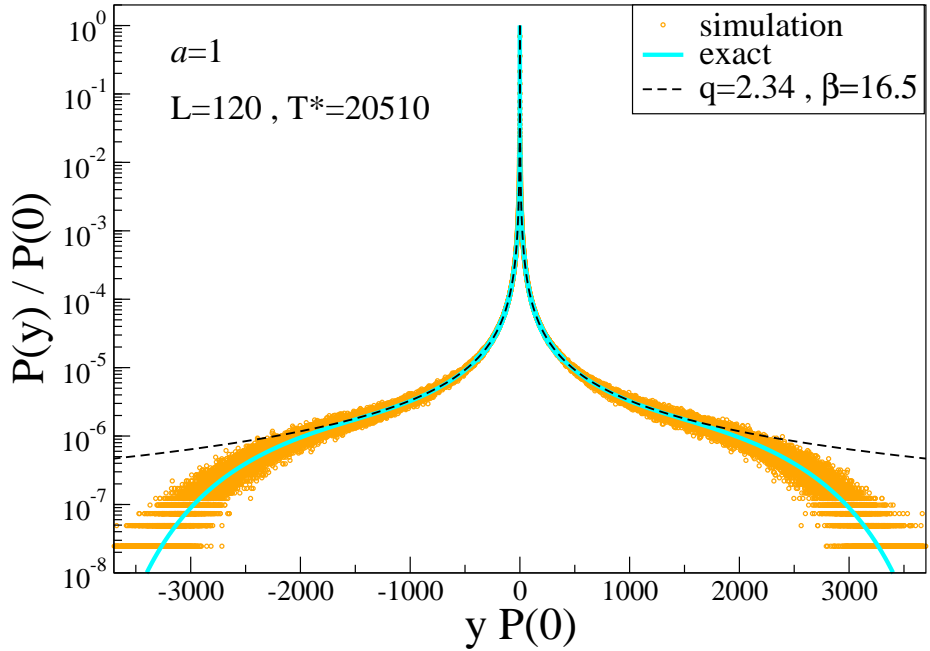}
\caption{Exact and simulation results of the case $a=1$, $L= 120$ and 
$p=5\cdot 10^{-6}$. 
It is clearly seen that the probability function $P(y,T^*)$ obtained from simulations 
is completely in accordance with the exact results. The number of experiments used in our simulations are $2x10^{8}$.
}
\label{Py-sim}
\end{figure}

\begin{figure}
\includegraphics*[scale=0.5]{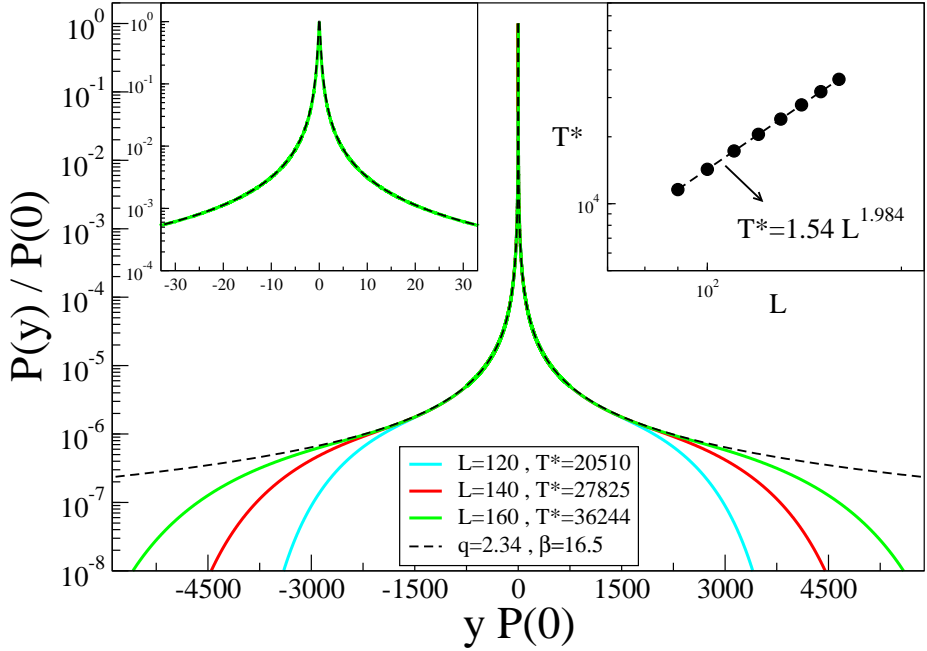}
\caption{The case $a=1$ for $L= 120$, $140$ and $160$ with $p=5\cdot 10^{-6}$. 
The main panel shows the probability 
function $P(y,T^*)$. The center of the function is shown in detail in the left inset. 
The time $T^*$ is chosen to optimize the fit to the $q$-Gaussian. The scaling of $T^*$ is 
given in the right inset.}
\label{Py}
\end{figure}

We now investigate the functional shape of the distribution $P(y,T)=\sum_xP(y,x,T)$ for 
different values of $a$, $L$ and $T$, for the typical value $p=5\cdot 10^{-6}$. 
In Fig.~\ref{Py-sim} we plot a typical case from where the perfect agreement 
between the exact and simulation results is evident. 
Fig.~\ref{Py} is concerned with the case $a=1$ 
for different values of $L$. For fixed $L$ we determine the value $T^*$ for which an 
optimal fit to a $q$-Gaussian is possible. 
For $T>T^*$ values, each curve will start to 
exceed the $q$-Gaussian tails before the fast drop region, due to the finite-size, 
is being achieved. 
It is also interesting to analyze the relationship between $T^*$ and $L$. We find 
that $T^*\sim L^2$, a behavior {\em identical} to the ordinary scaling that 
relates time and distance for diffusive processes.  

\begin{figure}
\includegraphics*[scale=0.5]{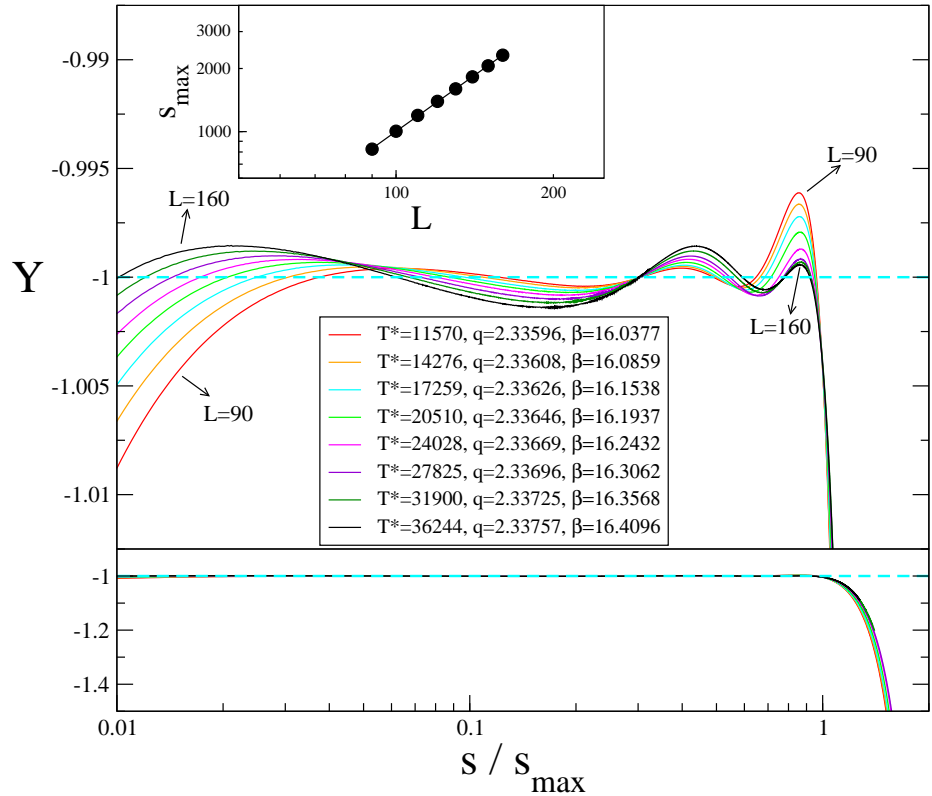}
\caption{The case $a=1$ and $p=5\cdot 10^{-6}$ with $L$ values between 
$L= 90$ to $160$. The lower panel shows the data collapse when the $x$-axis 
is appropriately scaled. 
The upper panel shows a zoomed region around $Y=-1$ line. 
In the inset the scaling of $s_{max}$ with $L$ is given. The straight line is $s_{max}=AL^C$ with $A=0.2726$ and $C=1.7828$.}
\label{diff_L}
\end{figure}

Since we have an exact numerical solution we can investigate with great accuracy 
the nature of the convergence to the $q$-Gaussian as we increase the domain $L$ of the random walker. 
In Fig.~\ref{diff_L} we demonstrate that as $L$ is increased a trajectory in the 
$T^*$-$q$-$\beta$ parameter space exists along which $P(y,T^*,L)$ becomes increasingly well 
described by a $q$-Gaussian. The Figure contains the scaling combination 
$Y\equiv\ln_q[P(y,T)/P(0,T)]/[\beta s^2]$, where $s=y P(0)$, 
$\ln_q(x)=(x^{1-q}-1)/(1-q)$ is the $q$-logarithm and the scaling parameter $s_{max}$ 
is defined as the $s$ value of each $L$ for which $Y$ significantly starts to deviate from the $-1$ line, namely when $|Y+1|>0.004$.  
If the dependence on $y$ is exactly $q$-Gaussian, we would have $Y=-1$ for 
all $y$. An appropriate scaling of $x$-axis yields a clear data collapse. 
For all values of $L$ we observe the deviation from $-1$ to be no more than a few 
parts in a 1000 and as $L$ increases the curves indeed approaches the line $Y=-1$ for large 
values of the argument $y$. The oscillations about the $Y=-1$ curve exhibit a subtle dependence on $L$. Careful inspection of the top panel in Fig. \ref{diff_L} reveals that for increasing values of $L$ the curves actually approach the $Y=-1$ line for both small and large values of the argument $s/s_{max}$. We therefore believe that asymptotically the distribution $P(y,T^*)$ indeed becomes very well described by the q-Gaussian functional form. It is unfortunately numerically impossible for us to reach very large $L$-values.

\begin{figure}
\includegraphics*[scale=0.5]{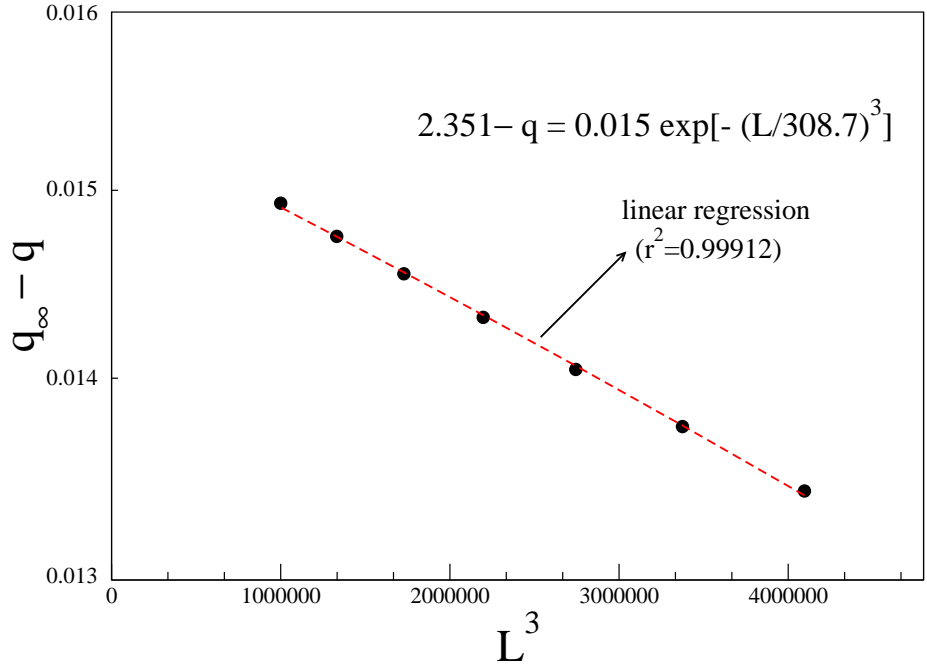}
\caption{Linear-log representation for the $L$ dependence of $q$ values. This exponential 
dependence suggests an asymptotic value around $q_{\infty}\simeq 2.351$.}
\label{qL}
\end{figure}

 This suggests the distribution $P(y,T,L)$ asymptotically is 
described by the $q$-Gaussian form if one let the pair $(L,T)$ vary appropriately. 
We localize these very precise $q$ and $\beta$ values so that the curves are as 
symmetric as possible along the $-1$ line.  
Using the $q$ values given in Fig.~\ref{diff_L}, we obtain an exponential dependence on 
$L$ from where one can predict the asymptotic value 
$\lim_{L\rightarrow \infty} q(L)=q_{\infty}\simeq 2.351$, which is evident from 
Fig.~\ref{qL}.  It is interesting to note that for values $5/3<q<3$ the variance diverges. So in this respect the distribution behaves similarly to e.g. the  Cauchy-Lorentz distribution, which corresponds to $q=2$. Diverging variance is of course a common feature  in complex systems of distributions with power law tails.

\begin{figure}
\includegraphics*[scale=0.5]{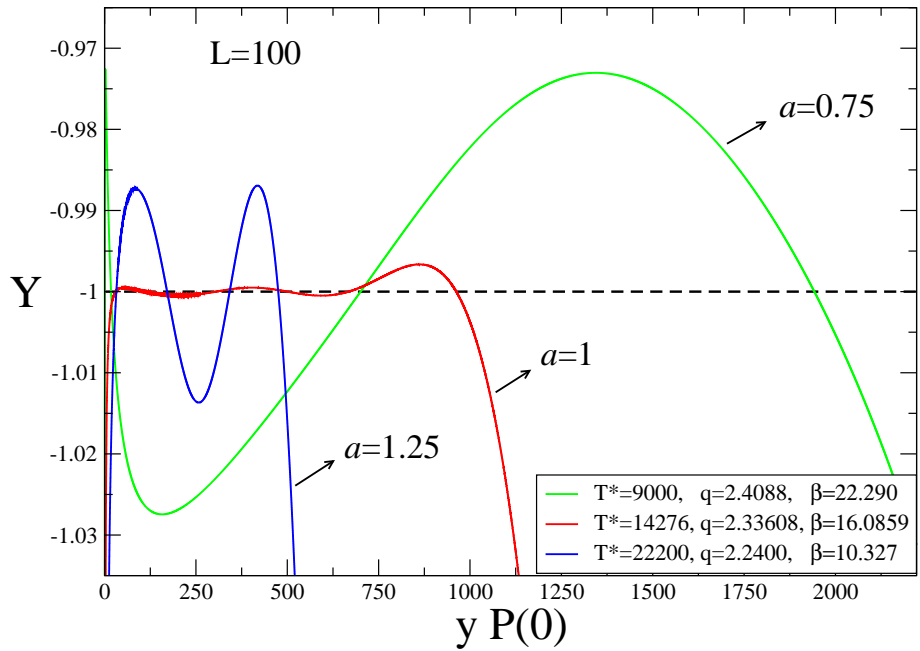}
\caption{$Y$ plot of cases $a=0.75$, $a=1$ and $a=1.25$ for a representative $L$ value. 
Whenever $a\neq 1$, increasing order of deviation from $-1$ line is evident.}
\label{a_diff_1}
\end{figure}

Next we consider the effect of changing $a$ to values different from one. 
Although it is still possible to tune $T^*$ so that the 
distribution $P(y,T^*)$ is very close to a  $q$-Gaussian, the high resolution 
$Y$ plot given in Fig.~\ref{a_diff_1} now shows that the order of deviation from 
straight horizontal line through $Y=-1$ grows significantly whenever $a\neq 1$.  

\section{The r{\^ o}le of correlations}
\begin{figure*}
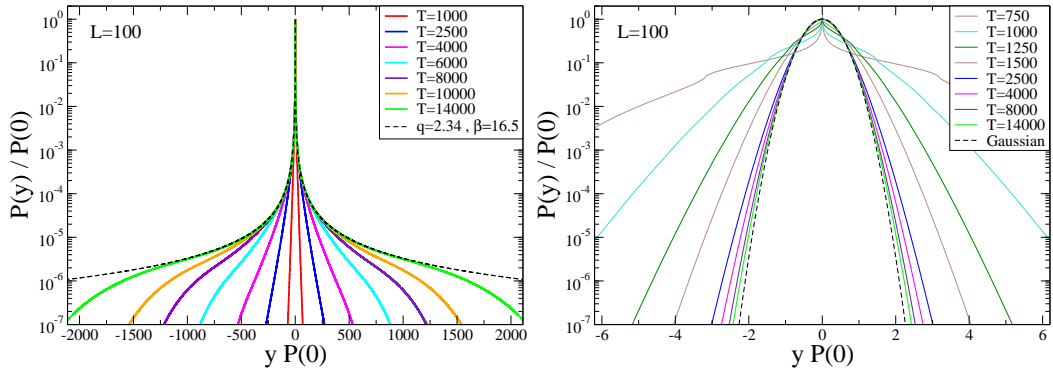

\includegraphics*[height=4.85cm]{fig7a.eps}
\includegraphics*[height=4.85cm]{fig7b.eps}
\caption{Comparison of the $T$ dependence for fixed values of $L$ with 
$p=5\cdot 10^{-6}$ for the correlated (left panel) and 
the uncorrelated (right panel) $P(y,T)$ distribution 
(see text for the details).  }
\label{uncorr}
\end{figure*}

One might perhaps wonder to what extent the observed 
deviation from ordinary Gaussian behavior is caused by the peculiar shape of the probability 
distribution of the individual terms $x_t$ in the sum  $y=\sum_{t=1}^Tx_t$. To check 
this we solved the Master Equation for the probability distribution for $y$ in the 
uncorrelated case where all the individual terms in the sum are drawn independently with 
probability
\begin{equation}
p_{uc}(x)=\frac{\cal N}{|\frac{x}{L}|+p},
\end{equation}
for $x\in\{-L,-L+1,..,0,..,L\}$ and $\cal N$ the normalization factor. The motivation for 
this is simply that for the RRW considered above a term will appear in the sum $y$ a 
number of times roughly given by $1/g(x)$. In Fig. \ref{uncorr} we show that when the terms 
are uncorrelated the sum converges towards an ordinary Gaussian. We note that the 
uncorrelated distribution $P(y,T)$ for small values of $T$ does resemble a $q$-Gaussian 
in the region of small $y$ values. However, as $T$ is increased the functional form 
rapidly changes towards the ordinary Gaussian in stark contrast to the correlated case 
(left panel in Fig.~\ref{uncorr}) where the $P(y,T)$ grows towards the $q$-Gaussian as 
$T$ is increased up to very large values of $T$.  For the uncorrelated sum no trajectory 
$(L,T^*)$ which for $L\rightarrow \infty$ takes one to the $q$-Gaussian exists. 


\section{Conclusions}
We have presented the hitherto most simple setting in which $q$-Gaussians 
control asymptotic behavior. We conclude that the $q$-Gaussian behavior is brought about by 
the strong correlations and the high reluctance for the walker to move away from the 
central region of its domain.

The numerical exact solution of Master Equations allows us to present high precision data 
for the probability function of sums of correlated random variables derived from a 
restricted random walk (RRW) with position dependent transition probabilities. When the 
range of the walker $L$ and the number of terms in the sum $T$ is scaled according to 
$T=1.54 L^2$, $q$-Gaussians are observed over an increasingly broad interval. 
For non-linear transition probabilities we are able to identify a subtle oscillatory 
behavior away from the pure $q$-Gaussian form. Given the relative simplicity of the RRW 
it appears likely that the relation between transition probability and the value of $q$ 
and the existence of oscillatory corrections to the $q$-Gaussian asymptote can be 
unraveled analytically. 
The RRW model presented in the present letter promises this way to significantly increase 
our understanding of the mechanisms responsible for the often encountered $q$-Gaussians. 

The very weak dependence  of $q$ on $T^*(L)$ and the subtle oscillations in the $a=1$ 
case and the more essential oscillations present for $a\neq 1$ indicates that the true exact 
mathematical asymptote might not strictly be $q$-Gaussian but rather some functional form 
resembling a $q$-Gaussian to a high degree of accuracy. It is only because we have 
numerically iterated the Master Equations exactly that we are able to identify this very slight 
difference. In studies relying on simulations or observational data the accuracy may not 
be sufficient to resolve these details and one would conclude that a $q$-Gaussian is an 
excellent approximation of the observed behavior. 

Let us recall that  $q$-Gaussians have been found previously for more complex processes than the random walk to be able to provide very high quality approximations to relevant distributions. The case  $q<1$ was considered in  \cite{moyano,thistleton} where the authors found for  two (scale-invariant) probabilistic models that the large-size limiting distributions are amazingly close to $q$-Gaussians, but are not exactly $q$-Gaussians \cite{hilhorst}. The work in Ref.  \cite{rodriguezstefan} provides an analytic example of large-size limiting distributions that are $q$-Gaussians.  We stress that even if $q$-Gaussians are not always the {\em exact} analytic form of the probability distributions in question, it is highly intriguing why they provide such exceptionally high accuracy approximations in a large number of cases where correlations are sufficiently strong to make the central limit theorem inapplicable.

\acknowledgments
H.J.J. and C.T. warmly thank the organizers of the
2nd Greek-Turkish Conference on Statistical Mechanics and Dynamical
Systems (Marmaris-Rhodos, 5-12 September 2010), where this work was initiated. 
H.J.J. is grateful for discussions with Gunnar Pruessner. 
The numerical calculations were performed in part at TUBITAK ULAKBIM, 
High Performance and Grid Computing Center (TR-Grid e-Infrastructure). 
This work has been supported by Ege University under the Research Project number 2009FEN077. 
Partial support by CNPq and Faperj (Brazilian agencies) is acknowledged as well.

\end{document}